\documentclass[useAMS,usenatbib]{mn2e}

\usepackage{amsmath}
\usepackage{epsfig}
\usepackage{xspace}

\newcommand{\ea}{et~al.\xspace}
\newcommand{\eg}{e.g.\xspace}
\newcommand{\pr}{\textrm{Pr}}
\newcommand{\Data}{\textrm{\,Data\,}}

\newcommand{\bmodel}{\textrm{\,Beta\,}}
\newcommand{\imodel}{\textrm{\,iHSE\,}}
\newcommand{\datav}{\mathbf{d}}

\newcommand{\parsv}{\bmath{\theta}}
\newcommand{\datum}{\mathnormal{d}}
\newcommand{\model}{\mathnormal{H}}
\newcommand{\param}{\theta}
\newcommand{\covmat}{\mathsf{C}}

\newcommand{\Om}{$\Omega_{\rm m}$}
\newcommand{\ob}{\Omega_{\rm b}}
\newcommand{\om}{\Omega_{\rm m}}

\title
[Bayesian joint analysis of cluster data]
{Bayesian joint analysis of cluster weak lensing and 
Sunyaev-Zel'dovich effect data}

\author[P.J.~Marshall~\ea]
{{P.J.\ Marshall},\thanks{Email: P.Marshall@mrao.cam.ac.uk}
 {M.P.\ Hobson}, 
 {A.\ Slosar} 
\\
Astrophysics Group, Cavendish Laboratory, Madingley Road, Cambridge
CB3 0HE
\\
}

\date{Accepted ???. Received ???; in original form \today}

\begin{document}
%
%
\maketitle
\begin{abstract}

As the quality of the available galaxy cluster data improves,  the
models fitted to these data might be expected to become increasingly
complex. Here we present the Bayesian approach to the problem of
cluster data modelling: starting from simple, physically motivated
parameterised functions to describe the cluster's gas density,
gravitational potential and temperature, we explore the
high-dimensional parameter spaces with a Markov-Chain Monte-Carlo
sampler, and compute 
the Bayesian evidence in order to make probabilistic statements about 
the models tested. In this way sufficiently good data will enable the 
models to be distinguished, enhancing our astrophysical understanding;
in any case the models may be marginalised over in the correct way when
estimating global, perhaps cosmological, parameters.  In this work we
apply this methodology to two sets of simulated interferometric
Sunyaev-Zel'dovich effect and  gravitational weak lensing data,
corresponding to current and next-generation telescopes. We calculate
the expected precision on the measurement of the cluster gas fraction
from such experiments, and investigate the effect of the primordial CMB
fluctuations on their accuracy. We find that data from instruments such
as AMI, when combined with wide-field ground-based weak lensing data,
should allow both  cluster model selection and estimation of gas
fractions to a precision of better than 30 percent for a given cluster.
 
\end{abstract}

\begin{keywords} 
methods: data analysis -- 
cosmology:observations  --
galaxies: clusters: general --  
cosmic microwave background -- 
cosmology: theory -- dark matter -- gravitational lensing
\end{keywords}


\section{Introduction}
\label{sect:intro}

Clusters of galaxies, as the largest gravitationally bound structures
in the Universe, may be used as cosmological probes. The number count
of clusters as a function of their mass has been predicted both
analytically~\citep[see \eg][]{COS/P+S74,COS/SMT01} and from large scale numerical 
simulations~\citep[see \eg][]{COS/Jen++01,COS/Evr++02}, and are very sensitive to
the cosmological parameters $\sigma_8$ and~\Om~\citep{COS/B+W03}. The
size and formation history of massive clusters is such that the ratio of gas mass to total
mass is expected to be representative of the universal ratio $\ob/\om$,
once the relatively  small amount of baryonic matter in the cluster
galaxies is taken into account~\citep{COS/Whi++93}. 

The deep gravitational potential wells of clusters contain hot ionised
gas, which radiates via thermal bremsstrahlung in the X-ray waveband,
with a luminosity proportional to the projected squared gas density. 
Inverse Compton scattering of cosmic microwave background radiation
(CMB) photons by this gas produces an observable shift in the thermal
spectrum of the CMB in the direction of the 
cluster~\citep[the Sunyaev-Zel'dovich (SZ) effect, ][]{SZ/S+Z72} 
which is
proportional to the projected gas density. Comparing the SZ and X-ray
flux densities allows the angular size distance, and so the Hubble
constant, to be
measured~\citep*{J/CDZ78}.

To date, most of our knowledge about clusters has come from optical
studies of the cluster member galaxies, and from X-ray observations of 
the intra-cluster medium. In particular, the recently  acquired data
from the Chandra and XMM missions have allowed the spatial and spectral
features of the X-ray emission from galaxy clusters to be  measured in
unprecedented detail.  
This has allowed the measurement of the mass function and global 
cluster gas fraction to be attempted~\citep{COS/ASF02,COS/All++03};
however, the methodology requires X-ray data from the most massive,
relaxed clusters, and includes assumptions of high symmetry and
hydrostatic equilibrium.

It is becoming increasingly common to measure the distribution
of gas and mass in galaxy clusters by other means. Weakly 
gravitationally lensed images of field 
galaxies lying behind the cluster are now routinely observed and used to
infer the projected mass distribution of the 
cluster~\citep[see \eg][]{GL/K+S93,GL/Dah++02,GL/C+S02}. Similarly, the
SZ effect has been observed in many
clusters~\citep[see \eg][]{SZ/Jon++93,SZ/CJL96,SZ/Gre++01}, and has
been used with some success in the determination of
H$_0$~\citep[see \eg][]{J/B+H94,J/Jon++01,J/MMR01,J/Gra++02,J/Ree++02}. 
Clearly a more complete 
analysis
of clusters of galaxies utilising all these data at the same time would
be desirable. In this paper we outline our general approach to this
task, and show how our methods give natural answers to the specific
questions being asked of the data. 
Either Hubble's
parameter, or the cluster gas fraction, or indeed both,
could equally well
be chosen as the cosmological parameter for the demonstrative purposes
of this work; for simplicity we focus on the gas fraction, setting 
H$_0 = 100 h\;\textrm{km s}^{-1}$. 

The quality of X-ray data in the past, and that of current weak lensing
and SZ data, is such that the information content of each dataset
individually matches that of a
simple, symmetric, constrained model, with just a handful of parameters.
One might expect that combining the datasets would lead to an
improvement in the parameter estimates, and thus open up the possibility
of testing more complex models. 
In this work we simulate and compare observations made with current
generation SZ telescopes, such as the 
VSA~(Watson~\ea~\citeyear{SZ/Wat++03}) and
CBI~(Pearson~\ea~\citeyear{COS/Pea++03}), 
with those under construction such as 
AMI~(Kneissl~\ea~\citeyear{SZ/Kne++01}) and the 
SZA~(Mohr~\ea~\citeyear{SZ/Moh++02}). 
In each case we match the mock SZ
observations with simulated wide field lensing data, and investigate the
prospects for measuring the cluster gas fraction via this route. 

In Section~\ref{sect:infer} we discuss a systematic
approach to the problem of cluster model parameter fitting and model
selection. We also show in this context how to infer model-independent
conclusions from cluster data, where the derived errors on, in
particular, any inferred cosmological parameters take the uncertainty 
in the cluster model into account. After an introduction to the example
cluster data and
models in Sections~\ref{sect:data} and~\ref{sect:models},  
we apply these
methods to simulated sets of data from 
weak gravitational lensing and the SZ effect
observed by microwave interferometers in Section~\ref{sect:sim}, 
and discuss the prospects of the aforementioned current and near-future
experiments. 
We then put the results of this demonstration 
in context with a 
brief discussion of our methodology and others 
(Section~\ref{sect:disc}), and summarise our
conclusions in Section~\ref{sect:concl}.


\section{Bayesian inference}
\label{sect:infer}

For a set of data arranged into a vector $\datav$, our knowledge of
the experimental errors on those data can be written in the form of a
likelihood function $L$:
\begin{equation}
L = \pr(\datav|\parsv,\model_j),
\end{equation}
which is a function of the observed data and depends on the parameter
vector~$\parsv$
of an assumed model or hypothesis~$\model_j$. 
By the central limit theorem
the likelihood can often be
approximated by a multivariate Gaussian, 
\begin{equation}
L = \frac{1}{|2\pi\covmat|^{\frac{1}{2}}} \exp{-\frac{1}{2}(\datav - \datav_p)^T\covmat^{-1}(\datav - \datav_p)}
\end{equation}
where the predicted data vector $\datav_p$ is calculated within the
model~$\model_j$. However, other forms of the likelihood function may
describe the data more accurately; none of the rest of this section
relies on the Gaussian approximation. If the data vector can be written
as
\begin{equation}
\label{eq:ind_data}
\datav = \datav_1 + \datav_2,
\end{equation}
where the two sub-datasets are independent, then we can write the
joint likelihood as
\begin{equation}
\pr(\datav|\parsv,\model_j) =
\pr(\datav_1|\parsv,\model_j)\pr(\datav_2|\parsv,\model_j),
\end{equation}
as has been applied to cluster potential analysis by~\citet{J/Cas++00}.

\noindent At this point we can ask the following questions:
\begin{enumerate}
\item ``What are the relative probabilities of any two models $\model_i$ and
$\model_j$ being the true model, given all the 
information we have?''
\item ``What is the joint probability distribution of the parameters
$\parsv$ of model $\model_i$, given all the 
information we have?''
\item ``What is the probability distribution of any one particularly
interesting parameter $\theta_k$, given all the 
information we have?''
\end{enumerate}
In the context of galaxy cluster data analysis, a model 
will be a set of suitably 
parameterised functions describing the cluster potential, gas density
and temperature, and a suitably
parameterised background cosmology; all of these parameters are included
in the vector $\parsv$ and should be
inferred simultaneously. Learning the structure of the cluster 
involves finding the most appropriate \emph{cluster} model and
its parameters; the first two questions above can therefore be
bracketed together as ``astrophysical'' questions. To do cosmology with
clusters one just wants to
investigate the cosmological part of the parameter space, in a
way that is independent of the cluster model. In this respect,
question~(iii) can be labelled ``cosmological''.

\subsection{Model selection and parameter fitting: astrophysics}

The answer to question~(ii) above is given by Bayes' theorem:
\begin{equation}
\label{eq:bayes1}
\pr(\parsv|\datav,\model_j) =
\frac{\pr(\datav|\parsv,\model_j)\pr(\parsv|\model_j)}{\pr(\datav|\model_j)}.
\end{equation}
The probability  density function (pdf) $\pr(\parsv|\model_j)$ should encode any 
prior knowledge we have of the parameters of the model in question. For
example, uncertainty as to the order of magnitude of a parameter should
be represented by a uniform probability distribution in the logarithm of
the parameter~\citep{ST/Jef32}, whilst a previously performed, 
independent measurement of a parameter might be interpreted as a 
Gaussian prior centred on the observed value with width equal to the 
quoted error. 

The denominator of equation~(\ref{eq:bayes1}) plays an important role in
answering question~(i). Applying Bayes' theorem again we have
\begin{equation}
\label{eq:bayes2}
\pr(\model_j|\datav) =
\frac{\pr(\datav|\model_j)\pr(\model_j)}{\pr(\datav)}.
\end{equation}
Here, the denominator is a constant. Moreover, if, prior to the 
analysis, the
various models are assigned equal probabilities (as might be the case
for an attempt at a ``fair test'', or if the analyst really does have
no more belief in any one model than another), then the quantity 
$\pr(\datav|\model_j)$ may be used in comparing the relative 
probailities of each model under consideration. For more details on the
use of this quantity, known as the 
``evidence''~\citep[see \eg][]{Jaynes,ST/Mac92}.
Calculation of the evidence is in principle straightforward;
marginalising out the $N_i$~parameters of the model gives
\begin{equation}
\pr(\datav|\model_j) = \int
\pr(\datav|\parsv,\model_j)\pr(\parsv|\model_j)\, d^{N_i}\parsv.
\label{eq:evint}
\end{equation} 
However, for the purposes of this work it is useful to note that the
evidence is sensitive both to the parameter prior and the
likelihood. 
Models providing a better fit to the data, and 
so a higher value at the peak of the likelihood function, have higher
evidence associated with them. Conversely, models whose priors
define large volumes of low likelihood parameter space will give lower
evidence values -- overly complex models are penalised.

We now turn to the practical problems associated with the calculation of
the posterior probability $\pr(\parsv|\datav,\model_j)$ and the
associated evidence~$\pr(\datav|\model_j)$. One approach, much used in
the field of cosmological parameter estimation, is to calculate
the prior density and likelihood on a discrete parameter grid. However,
as the models considered become more complex, this task becomes
exponentially more computationally intensive. As an illustration,
if a likelihood evaluation for a 10 parameter model takes just 1 second,
the time to produce a grid of posterior probability densities, ten pixels
in each dimension, 
is of order $10^{10}$ seconds, or 300 years. Clearly this procedure is
neither practical nor accurate. Instead, we use an exploratory 
Markov-Chain Monte-Carlo (MCMC) algorithm to draw samples from the 
multidimensional unnormalised posterior 
$\pr(\datav|\parsv,\model_j)\pr(\parsv|\model_j)$~\citep{MCMC}.
The output of the MCMC algorithm is a list of samples whose number 
density in
parameter space is proportional to the posterior probability density. 
This allows a
representation of the desired posterior to be constructed which is both
efficiently calculated (computation time is typically of the order of 
a few hours) and convenient to use. For example, the sample parameter 
values may be
combined to calculate predictions for other properties of the model.

Calculating the evidence by performing the 
integral of equation~(\ref{eq:evint}) 
numerically by ordinary means would 
of course involve a similar number of operations as that outlined above.
We instead make use of the technique of ``thermodynamic integration'' to
calculate the evidence dynamically during an initial ``burn-in'' phase
of the MCMC algorithm~\citep{MCMC}. 
This process results in evidence values precise to a fraction of a 
unit in
$\log_e{\pr(\datav|\model_j)}$. This corresponds to a probability ratio
between two models' of less than 3, well below the ``belief threshold''
of a sensible analyst.

\subsection{Marginalising over parameters and models: cosmology}

A major benefit of working with samples drawn from the posterior rather
than a grid of posterior values is that the process of marginalisation
becomes trivial. When estimating a single parameter~$\theta_i$ of a 
model~$\model_j$ the distribution of interest is
\begin{equation}
\pr(\param_i|\datav,\model_j) = \int
\pr(\parsv|\datav,\model_j)\, d^{{N_{i}}-1}\parsv,
\end{equation} 
where the integral is over all other parameters. This projection takes
into account all the parameter degeneracies of the model, as well as
the priors on all of the parameters. An ensemble of sample
$\parsv$-vectors drawn from the full posterior can be projected onto the
$\param_i$ direction just by extracting the $\param_i$ values from the
ensemble. 

Although computing $n$-point statistics from a set of samples is very
easy, reconstructing the marginalised posterior pdf is not so
straightforward. 
To do this we provide as estimators for these
distributions histograms of sample values 
smoothed to some arbitrary length
scale, such that we err on the side of caution and never
underestimate the distribution widths. 
This is in keeping with our general
approach: the marginalised posterior probability distribution 
$\pr(\param_k|\datav)$ is
in general broader than those conditional on other parameters being 
fixed at, for instance, the maximum likelihood point. The resulting
estimators are therefore as accurate as possible, since the maximum
amount of
information has been included, and as precise as allowed by the 
quality of
the data, since the errors on the observed quantities have been
rigorously propagated in a self-consistent way.

By taking this process one step further we can answer question~(iii), by
marginalising over model space. 
This procedure is important in the context of cluster data analysis: we
want to be able to make robust,
model-independent statements about cosmological parameters 
from cluster
data. To this end, and denoting the cosmological parameter of 
interest~$\param_k$,
one should calculate
\begin{align}
\label{eq:marg_models1}
\pr(\param_k|\datav) &=\underset{j}{\sum} \pr(\param_k|\datav,\model_j)\pr(\model_j|\datav) \\
\label{eq:marg_models2}
&\propto \underset{j}{\sum}
\pr(\param_k|\datav,\model_j)\pr(\datav|\model_j)\pr(\model_j).
\end{align}
The last proportionality is as such because we have discarded the 
normalising
constant of equation~(\ref{eq:bayes2}). Equation~(\ref{eq:marg_models2}) 
is now a relationship between
quantities that we can calculate, and represents a model-averaging
process. We might hope that one particular model would be many times
more probable given the data, in which case we have learnt something
about the astrophysics of the cluster and the sum will be
dominated by this model. On the other hand, it is always possible to
construct a range of models whose parameter priors all match the data's
likelihood equally well, and so
give similar evidences. In this case the averaging process serves
to increase the width of the posterior density of interest, that of the
interesting parameter $\param_k$ given the data only, by an amount
appropriate to the uncertainty over the range of models. In this way,
(quasi)-model independent statements can be made about cosmological
parameters, such as the cluster gas fraction, from the
cluster data.


\section{Data}
\label{sect:data}

In this section we give a brief introduction to weak lensing and
interferometric Sunyaev-Zel'dovich effect data, demonstrating the
construction of the likelihood function in each case.

\subsection{Weak lensing}
\label{subsect:data:lensing}

Weak gravitational lensing may be used to investigate cluster mass
distributions through the relationship~\citep[see \eg][]{GL/S+K95}
\begin{equation}
\langle \epsilon \rangle = g(\parsv);
\label{eq:e-g}
\end{equation} 
that is, the average complex 
ellipticity~$\epsilon = \epsilon_1 + i \epsilon_2$ 
of an ensemble of
background galaxy images is an unbiased estimator of
the local reduced shear field~($g$)
due to the cluster. 
Equation~(\ref{eq:e-g})
allows us to use each of the~$2N_{\rm gal}$
lensed ellipticity components~$\epsilon_j$ of $N_{\rm gal}$ measured 
background
galaxy images as noisy estimators for the corresponding component of the
reduced shear~$g_j(\parsv)$
due to the cluster model, parameterised 
with the variables~$\parsv$~\citep{GL/SSB98,GL/SKE00,GL/Mar++02}. 

Here, the complex ellipticity is defined such that an elliptical object
with semi-major axis~$a$, semi-minor axis~$b$ and orientation angle
(measured anticlockwise from $x$-axis to semi-major axis)~$\phi$ has
ellipticity
\begin{equation}
\epsilon  = \frac{a-b}{a+b}e^{2 i \phi}.
\label{eq:edef}
\end{equation} 
Figure~1 of~\citet{GL/KSB95} shows the image shapes corresponding to
different values of this (or a similarly defined) ellipticity; the 
effect
of an isolated mass concentration lying in front of a galaxy field 
is to align the image shapes tangentially about the centre of the mass
distribution.
Large catalogues of background galaxies with measured ellipticities are
now almost routinely generated from wide field optical images, using
shape estimation software such as~\texttt{imcat}~\citep{GL/KSB95} or
\texttt{im2shape}~\citep{GL/Bri++01}.

We may organise the observed ellipticities into the data vector $\datav$, 
having components
\begin{equation}
\datum_i = \left\{ 
\begin{array}{ll}
\textrm{Re}(\epsilon_{i}) & \mbox{$\left(i \leq N_{\rm gal}\right)$} \\
 & \\
\textrm{Im}(\epsilon_{i-N_{\rm gal}}) & \mbox{$\left(N_{\rm gal}+1 \leq i \leq 2N_{\rm gal}\right)$}
\end{array}
\right.
.
\label{eq:gldvect}
\end{equation}
Likewise the corresponding model reduced shears can be arranged into 
the predicted data vector $\datav^{\rm p}$, having components
\begin{equation}
\datum^{\rm p}_i = \left\{ 
\begin{array}{ll}
\textrm{Re}(g_{i}) & \mbox{$\left(i \leq N_{\rm gal}\right)$} \\
 & \\
\textrm{Im}(g_{i-N_{\rm gal}}) & \mbox{$\left(N_{\rm gal}+1 \leq i \leq 2N_{\rm gal}\right)$}
\end{array}
\right.
.
\label{eq:gldvectp}
\end{equation}

The unlensed ellipticity components
may often be taken as having been
drawn independently from
a Gaussian distribution with mean~$g_j$ and 
variance~$\sigma_{\rm intrinsic}^2$, leading to a diagonal noise
covariance matrix~$\mathsf{C}$.
We can then
write the likelihood function as 
\begin{align}
L_{\rm Lensing} &= \pr(\mbox{\mbox{Data}}|\parsv) \\
\notag
&= \frac{1}{Z_L} \exp ( -\frac{\chi^2}{2} ),
\label{eq:gllhood}
\end{align} 
where $\chi^2$ is the usual misfit statistic 
\begin{align}
\chi^2 &= \sum_{i=1}^{N_{\rm gal}} \sum_{j=1}^{2} \frac{\left(\epsilon_{j,i} -
g_{j,i}(\parsv)\right)^2}{\sigma^2} \\
&= \left( \datav - \datav^{\rm p} \right)^{\rm T} \mathsf{C}^{-1} \left( \datav -
\datav^{\rm p} \right),
\label{eq:glchisq}
\end{align} 
and the normalisation factor is
\begin{equation}
Z_L = (2 \pi)^{2N_{\rm gal}/2} |\mathsf{C}|^{1/2}.
\label{eq:glchisqnorm}
\end{equation} 

The effect of Gaussian errors introduced by the galaxy shape estimation procedure
has been included by adding them in quadrature to the intrinsic
ellipticity dispersion~\citep{GL/HFK00,GL/Mar++02}, 
\begin{equation}
\sigma = \sqrt{\sigma^2_{\rm obs}+\sigma^2_{\rm intrinsic}}.
\label{eq:newsigma}
\end{equation} 
This approximation
includes the assumption
that 
the applied reduced shear is not too large (as is the case for
low redshift clusters).

\subsection{Sunyaev Zel'dovich effect}
\label{subsect:data:sz}

To date, the majority of the observations of the SZ effect towards 
clusters of galaxies have been made with 
interferometers~\citep[see \eg][]{SZ/Joy++01,J/Jon++01,SZ/Lar++03}.
These instruments have a number of advantages over single dish
telescopes, including their relative insensitivity to atmospheric 
emission~\citep[\eg][]{SZ/L+H00}, lack of required receiver stability, and the ease with which
systematic errors such as ground 
spill~(Watson~\ea~\citeyear{SZ/Wat++03})
and point source
contamination~(Grainger~\ea~\citeyear{SZ/Gra++02b}, Taylor~\ea~\citeyear{COS/Tay++03})
can be minimised. 

Assuming a small field size, an interferometer operating at a single
frequency~$\nu$ measures samples from the complex 
visibility plane $\widetilde{I}_{\nu}(\bmath{u})$. 
This is given by the weighted
Fourier transform of the surface brightness~$I_{\nu}$,
\begin{equation}
\widetilde{I}_{\nu}(\bmath{u}) 
= \int A_{\nu}(\bmath{x}) I_{\nu}(\bmath{x}) \exp(2\upi i\bmath{u}\cdot\bmath{x})~{\rm d}^2\bmath{x},
\label{eq:visdef}
\end{equation}
where $\bmath{x}$ is the position relative to the phase centre,
$A(\bmath{x},\nu)$ is the (power) primary beam of the antennas at the
observing frequency $\nu$ (normalised to unity at its peak), and
$\bmath{u}$ is a baseline vector in units of wavelength. 

The positions in the $uv$-plane at which this function is sampled by
the interferometer are determined by the physical positions of its
antennas and the direction of the field on the sky. The samples
$\bmath{u}_j$ lie on a series of curves which we may denote by the function
$\widetilde{B_{\nu}}(\bmath{u})$ that equals unity where the Fourier
domain (or $uv$-plane) is sampled and equals zero elsewhere. The
function $\widetilde{B_{\nu}}(\bmath{u})$ may be inverse Fourier
transformed to give the synthesised beam $B_{\nu}(\bmath{x})$ of
the interferometer at an observing frequency $\nu$. 

For a realistic
interferometer, the sample values will also contain a contribution due
to noise; the
$j$th baseline $\bmath{u}_j$ of an interferometer measures the 
complex visibility
\begin{equation}
V(\bmath{u}_j) = \widetilde{I}_{\nu}(\bmath{u}_j) + N(\bmath{u}_j),
\label{eq:visobs}
\end{equation}
where $N(\bmath{u}_j)$ is the noise on the
$j$th visibility. This noise comes from two sources: the first is 
uncorrelated Johnson
noise from the receivers. 
The second source of noise is the CMB itself, now known to be
very well approximated by a Gaussian random field defined by its 
angular power spectrum. 
That is, expanding the primordial CMB
surface brightness distribution into a spherical harmonic series gives
\begin{equation}
I_{\nu}^{\rm CMB} = \sum_{\ell=0}^{\infty} \sum_{m=-\ell}^{\ell}
a_{\ell m}(\nu) Y_{\ell m} (\hat{\bmath{x}}),
\label{eq:cmbspec1}
\end{equation}
leading to the power spectrum
\begin{equation}
c_{\ell} = \langle |a_{\ell m}|^2\rangle.
\end{equation}
For small fields of view, the index~$\ell$ is related to the Fourier
($uv$) plane vector ~$\bmath{u}$ by $\ell\approx2\pi|\bmath{u}|$.
An ideal interferometer would therefore measure the~$a_{\ell m}$
directly, giving an extra noise vector 
\begin{equation}
N^{\rm CMB}(|\bmath{u}|) = \sqrt c_{\ell}.
\end{equation}
In practice, the interferometer observes the sky surface brightness 
multiplied by the primary beam~$A_{\nu}(\bmath{x})$; the corresponding
convolution with the aperture illumination 
function~$\widetilde{A}_{\nu}(\bmath{u})$ in the $uv$-plane 
produces correlated visibilities. Since
the SZ effect produces a (frequency dependent) linear 
perturbation~$\delta I_{\nu}$ to the CMB surface brightness, the
noise on an SZ observation can therefore be described by the covariance
matrix
\begin{equation}
\mathsf{C} = \mathsf{C}^{\rm receiver} + \mathsf{C}^{\rm CMB}.
\end{equation}
The first term on the right hand side is a diagonal matrix with 
elements $\sigma^2_i\delta_{ij}$, with $\sigma_i$ the rms Johnson noise
on the $i^{th}$ baseline visibility. The second term contains
significant off-diagonal elements and can be calculated
from a given primordial CMB power spectrum following~\citet{COS/H+M02}.
Rather than dealing with complex data and covariance matrices, it is
convenient to order the visibility components into a data
vector~$\datav$ with components 
\begin{equation}
\datum_i = \left\{ 
\begin{array}{ll}
\textrm{Re}(V_i) & \mbox{$\left(i \leq N_{\rm vis}\right)$} \\
 & \\
\textrm{Im}(V_{i-N_{\rm vis}}) & \mbox{$\left(N_{\rm vis}+1 \leq i \leq 2N_{\rm vis}\right)$}
\end{array}
\right.
. 
\label{eq:szdvect}
\end{equation}

\begin{figure}
\centering\epsfig{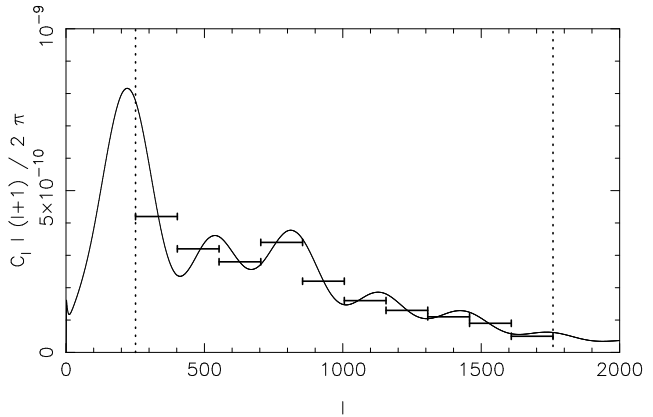}
\caption{ The flat band powers used to create the
covariance matrix associated with the primary CMB fluctuations on the
sky. The dotted lines correspond roughly to the range of scales that 
can be probed by the VSA in its present extended configuration. 
Overplotted is a
theoretical model corresponding to the flat $\Lambda$CDM model favoured by
the WMAP satellite data.}
\label{fig:cmbspec}
\end{figure}

With an inversionally symmetric primary beam pattern there is no
correlation between the real and imaginary parts of the visibilities, 
so with this
ordering of the data the matrix $\mathsf{C}^{\rm CMB}$ is
block-diagonal.  
The power spectrum is often
approximated to be constant within each of a set of bins in the
spherical harmonic coefficient index~$l$;
these bins therefore correspond to annuli in the $uv$-plane, and the
``flat band powers''~$d_b = \langle \ell(\ell+1) c_{\ell} \rangle_b$ 
are then related to the CMB covariance matrix
by an equation of the form 
\begin{equation}
\mathsf{C}^{\rm CMB}_{\mathnormal{ij}} = \sum_{b}
d_{b}
J_{\mathnormal{ij}}(|\bmath{u}|_b,|\bmath{u}|_{b+1}),
\label{eq:cldef}
\end{equation}
where the $b^{th}$ bin covers the range $|\bmath{u}|_b$ to
$|\bmath{u}|_{b+1}$, and the integrals $J_{ij}$ take the effect of the 
the aperture illumination function into account. 
Figure~\ref{fig:cmbspec} shows the power spectrum used in the
generation of the covariance matrix~$\mathsf{C}^{\rm CMB}$ for simulated
VSA data. 

With the combined covariance matrix~$\mathsf{C}$ in hand, the likelihood
function can be written
\begin{align}
L_{\rm SZ} &= \pr(\mbox{\mbox{Data}}|\parsv) \\
\notag
&= \frac{1}{Z_L} \exp \left( -\frac{\chi^2}{2} \right),
\label{eq:szlhood}
\end{align} 
where $\chi^2$ is again a statistic quantifying the misfit between
observed data~$\datav$ and predicted data~$\datav^p$, the latter of
which is a function of the model SZ surface brightness~$\delta I_{\nu}$:
\begin{equation}
\chi^2 = \left( \datav - \datav^{\rm p} \right)^{\rm T}
\mathsf{C}^{-1} \left( \datav - \datav^{\rm p} \right),
\label{eq:szchisq}
\end{equation} 
and the normalisation factor is
\begin{equation}
Z_L = \left(2 \pi \right)^{2N_{\rm vis}/2} |\mathsf{C}|^{1/2}.
\label{eq:szchisqnorm}
\end{equation}

\subsection{Joint analysis}
\label{subsect:data:joint}

As given in equation~(\ref{eq:ind_data}), the independent likelihoods  
described in
the previous subsections can be simply combined to give the joint
likelihood
\begin{equation}
\log{L} = \log{L_{\rm SZ}} + \log{L_{\rm Lensing}}.
\end{equation}
If the datasets contain systematic errors, then some attempt to deal
with this can be made by applying hyperparameters to the
likelihoods~\citep{COS/Lah++00,COS/HBL02}; this further
complication is not considered here. It is sufficient to note that such
a mismatch between the conclusions drawn independently from the two
datasets should result in a joint log evidence smaller than the sum of 
the two individual log evidence values.


\section{Simple cluster models}
\label{sect:models}
 
Clearly the sum over models in equation~(\ref{eq:marg_models2}) can be
simplified by discarding the terms with neglible weight; this
corresponds to investigating only physically reasonable models, starting
with the simplest and gradually increasing the complexity as required by
the data via the evidence. With this in mind, and for the illustrative purpose of this
work we limit ourselves to investigating just two cluster models,
referred to as ``Beta'' and ``iHSE''. 

Both models are spherically symmetric, with centroids assumed to be
known to within a Gaussian error of $\pm 1$~arcmin.  All projected
distributions are then circularly symmetric, with profiles defined as a
function of projected radius  $s = \sqrt{(x-x_0)^2 + (y-y_0)^2}$. The
three distributions taken to be fundamental in defining the cluster
model are taken to be the gravitational potential due to the total
mass density (including dark matter and intracluster medium), the
density of hot gas in the cluster, and the gas temperature. 

\subsection{Mass distributions}
\label{subsect:models:mass}
 
For simplicity, we limit ourselves in this  work to a  single model for
the mass distribution, and choose for this purpose the  NFW 
profile~\citep*{CS/NFW95}. This has been found to provide a good fit to
many numerically simulated clusters, and is simple enough that analytic
formulae for many derived distributions have been worked out, as
described below.
The three-dimensional radial dependence of
the density is given by 
\begin{equation}
\rho(r) = 
\frac{\rho_{s}}{\left(r/r_{s}\right)
\left(1+r/r_{s}\right)^{2}}.
\label{eq:nfwprofile}
\end{equation}
The gravitational lensing data are sensitive only to the projected
total mass
density distribution, $\Sigma(x)$, where $x$ is the 
scaled projected radius $x=s/r_s$. 
For a circularly symmetric surface density, 
the magnitude of the shear is given by
\begin{equation}
|\gamma| = \frac{\overline{\Sigma}(x) - \Sigma(x)}{\Sigma_{\rm crit}} 
\end{equation}
(with the overbar denoting average density within~$x$). 
Here, the critical density 
$\Sigma_{\rm crit}$ is a factor dependent on the angular diameter
distances to and between the lens~($l$) and source~($s$) planes:
\begin{equation}
\Sigma_{\rm crit} = \frac{c^2}{4 \pi G}\frac{D_s}{D_l D_{ls}}. 
\end{equation}
The complex reduced shear at background galaxy position 
radius~$x$ and azimuthal angle~$\phi$ 
can then be formed from
\begin{equation}
g = \frac{|\gamma|(x)}{1-\kappa(x)} \left( -\cos(2\phi) -i \sin(2\phi) \right),
\label{eq:nfwrshear}
\end{equation}
(with the convergence $\kappa=\Sigma/\Sigma_{\rm crit}$) 
and provide the predicted lensing data.
Analytical formulae for $|\gamma|(x)$ and $\kappa(x)$ are given
by~\citet{GL/W+B00} and are not reproduced here.

\begin{figure*}
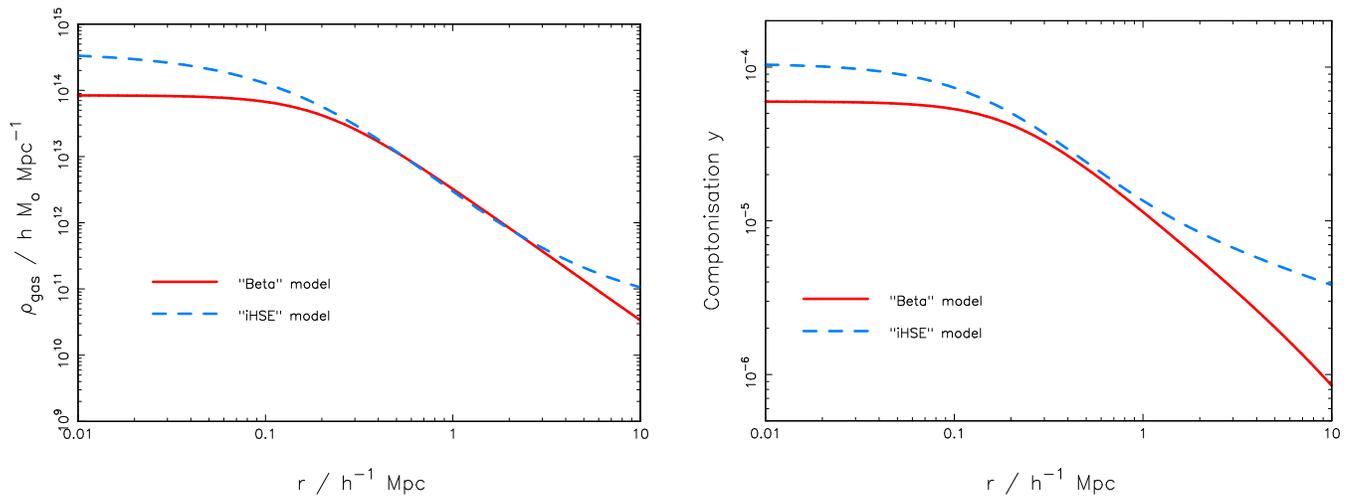

\begin{minipage}[b]{0.481\linewidth}
\centering\epsfig{file=rhogas.ps,width=\linewidth,angle=0,clip=}
\end{minipage} \hfill
\begin{minipage}[b]{0.479\linewidth}
\centering\epsfig{file=y.ps,width=\linewidth,angle=0,clip=}
\end{minipage}
\caption{ Deprojected radial gas density (left), and 
Comptonisation parameter~$y$ (right) for the two simulated clusters described 
in the text.}
\label{fig:profiles}
\end{figure*}
Note that for the NFW model 
the predicted lensing data are most easily parameterised in terms of a scale
radius and density; however, the prior which we understand best is on
the total mass, defined to be that within a radius~$r_{200}$ such that 
the average density within~$r_{200}$ is
200 times the critical density of the Universe. The overdensity 200
is a value often used by workers investigating numerical simulations
of clusters, a potential source of prior information. Given
values of $M_{200}$ and the corresponding concentration parameter
$c_{200}=r_{200}/r_s$, the scale radius and density can be
computed.

\subsection{Gas distributions}
\label{subsect:models:gas}
 
The SZ data are sensitive only to the cluster gas pressure, which must
be calculated from the model gas density and temperature distributions.
Both models assume a one-parameter isothermal temperature profile.
The difference between the models comes in the gas density distribution.
The Beta model has gas density profile
\begin{equation}
\rho_{\rm gas}(r) = \frac{\rho_{\rm gas}(0)}{\left(1+\left(r/r_{c}\right)^{2}\right)^{\frac{3\beta}{2}}},
\label{eq:betaprofile}
\end{equation}
often
used in cluster modelling~\citep[see \eg ][]{Sarazin}. In the iHSE model
the gas
is in full hydrostatic pressure equilibrium with the potential defined
by the NFW mass model (iHSE). This potential is
\begin{align}
\Phi &= -\frac{G M(r)}{r} \\
     &= -\frac{4 \pi G \rho_s r_s^3}{r} \left[ \log(1+r/r_s) -
     \frac{r/r_s}{1+r/r_s}
     \right].
\label{eq:nfwpotential}
\end{align}
The equation of hydrostatic equilibrium is 
\begin{equation}
\nabla P = -\rho_{\rm gas} \nabla \Phi,
\label{eq:hse}
\end{equation}
which, with the assumption of a spherically symmetric distribution of
ideal gas at temperature~$T$, becomes
\begin{equation}
\frac{d \log{\rho_{\rm gas}}}{d \log{r}} = -\frac{G M(r) \mu}{k T r}.
\end{equation}
We assume a mass per particle of $\mu = 0.59$ times the proton mass.
Given the potential of equation~(\ref{eq:nfwpotential}), this equation 
can be
integrated~\citep{CS/SSM98} to give
\begin{equation}
\rho_{\rm gas} = \rho_{\rm gas}(0) \exp{-\frac{4 \pi G \rho_s r_s^2
\mu}{k T}\left( 1- \frac{\log(1+r/r_s)}{r/r_s} \right)}
\label{eq:gasdensity}
\end{equation}
  
Note again that the predicted data are most easily expressed in terms of
a central gas density $\rho_{gas}(0)$, which can be computed by
numerically integrating either gas density profile to $r_{200}$ and
normalising to the gas mass within this radius, a parameter for which
again the prior is better understood. This integral is performed
numerically, as is the Abel integral used in projecting the gas pressure
distribution along the line of sight~$l$, as 
required in the calculation of the Compton $y$-parameter:
\begin{align}
y(s) &= \frac{\sigma_{\rm T}}{m_e c^2} \int_{-\infty}^{\infty} n_e k T dl \\
  &\propto \int_{s}^{\infty} \frac{2r \rho_{\rm gas}(r)}{\sqrt{r^2 -
  s^2}} dr.
\label{eq:ypar}
\end{align}
Predicted visibilities are generated by sampling the (Fast) Fourier
transform of the 30~GHz sky surface brightness~$\delta I_{\nu}$, related to the
$y$-parameter by
\begin{equation}
\delta I_{\nu} = f(\nu) y B_{\nu}(T_{\rm CMB}),
\end{equation}
where $B_{\nu}(T_{\rm CMB})$ is the CMB blackbody spectrum and $f(\nu)$
is a frequency-dependent factor approximately 
equal to $-2$ at 30~GHz.

Figure~\ref{fig:profiles} shows the Beta and iHSE model profiles, 
as functions of three-dimensional radius for the gas density
profile (equation~(\ref{eq:gasdensity})), and of projected radius for the Comptonisation
parameter (equation~(\ref{eq:ypar})).
Finally, the reader should note that the Beta profile, free from the
hydrostatic equilibrium constraint, has two extra parameters, making
it more flexible in fitting the data. Moreover, the two gas density
profiles described here have been chosen to have the same method of
normalisation, allowing straightforward comparison of the two models in
fitting the data. This is purely a matter of convenience, and is not
required by the methodology of Section~\ref{sect:infer}.
 

\section{Application to simulated data}
\label{sect:sim}

\begin{table*}
\caption{Cluster model parameters and priors. Inequalities denote
uniform prior probability between the given limits, whilst $(a \pm b)$
indicates a Gaussian prior centred on~$a$ with variance~$b^2$.}
\label{tab:models} 
\begin{tabular}{ccccc}
Dataset		& Model	& Parameters& Truth	& Priors				\\ \hline\hline 
Both			&		& $x_0,y_0$	& 0.0,0.0	& $(0 \pm 1)$~arcmin  \\ \hline
Gravitational	& Both	& $M_{200}$	& $7.0\times10^{14} h^{-1} \textrm{M}_{\odot}$ & $0 < M_{200} / h^{-1} \textrm{M}_{\odot} < 2\times10^{15}$ \\
lensing		&		& $c_{200}$	& 4 		& $0 < c_{200} < 12$                               \\ \hline
SZ effect		& Both	& $M_{\rm gas}$ & $4.9\times10^{13} h^{-2} \textrm{M}_{\odot}$ & $0 < M_{\rm gas} / h^{-2} \textrm{M}_{\odot} < 3\times10^{14}$ \\
			&		& $T$       & 8 keV 	& $(8 \pm 2)$~keV                         \\
			& Beta	& $r_c$     & $200 h^{-1}$~kpc	& $0 < r_c / h^{-1} \textrm{kpc} < 1000$             \\
			& only	& $\beta$   & 0.667 	& $0.3 < \beta < 1.5$                             \\ \hline
\end{tabular}
\end{table*}
In this section we use simulated data to demonstrate the methods
outlined in Section~\ref{sect:infer}. We first focus on simulations of 
current data,
namely observations of low redshift clusters in the optical with MegaCam
at CFHT~(Marshall~\ea 2003 in preparation), and at 30~GHz with the
extended
VSA~(Lancaster~\ea 2003 in preparation). 
We then move on
to consider data of the quality we might expect in the near future, at
higher redshift with SZ telescopes such as AMI. This is considered in 
combination with ground-based lensing data from a camera with
field of view again well-matched to the SZ
observations.

Our strategy is to simulate lensing and SZ 
data using each of the model clusters described in the previous section,
and then
analyse these data as recommended in Section~\ref{sect:infer}, assuming
in turn both of the models (one of which is the correct one). The
``true'' cluster parameters for each model are given in
Table~\ref{tab:models}. We use the same model clusters for both current
and next-generation mock observations, the only difference being a shift
in redshift from 0.07 (current observations) to 0.2 
(next-generation observations). Also given in 
Table~\ref{tab:models} are the prior pdfs used in the mock analyses.
The profiles plotted in Figure~\ref{fig:profiles} correspond to these
choices of true parameters.

The lensing data were generated by drawing $N$ background galaxy
ellipticities from a Gaussian intrinsic ellipticity distribution of
width 0.25, lensing them by the calculated reduced shear field of the
cluster, then adding realistic Gaussian shape measurement noise with 
$\sigma_{\rm obs}=0.2$. $N$~was specified by choosing a source density
of 15~arcmin$^{-2}$, appropriate to a 3-hour ground-based
observation~\citep{GL/C+S02}.

For the mock SZ datasets, the Fourier transform
of the model cluster's surface brightness was calculated at each of the
telescope-sampled
points in the $uv$-plane; Gaussian noise was then added, drawn from the
covariance matrix described in
Section~\ref{subsect:data:sz} with thermal receiver noise corresponding
to 150 hours' observation.


\subsection{Measuring cluster gas fractions}
\label{subsect:sim:fgas}

With the model gas and total mass density profiles both normalised to
the respective mass within~$r_{200}$, it is straightforward to compute
the cluster gas fraction within the same radius. Since the observable
quantities, the Comptonisation parameter~$y$ and the reduced shear~$g$,
cannot depend on the Hubble constant, the gas mass must have units of 
$h^{-2} \textrm{M}_{\odot}$ while the units of total mass are
$h^{-1} \textrm{M}_{\odot}$. Consequently, the gas fraction measurable
by combination of weak lensing and SZ effect data contains a factor 
of~$h$:
\begin{equation}
\frac{M_{\rm gas}}{M_{200}} = f_{\rm gas} h.
\end{equation}
This is the same $h$-dependence as found by~\citet{SZ/Gre++01} when
constraining the potential with the assumption of hydrostatic
equilibrium, but is different from that arising from the fitting of
X-ray data. Combination of the results from lensing and SZ data with
those from an analysis of X-ray data will thus break the degeneracy
between the gas fraction and the Hubble parameter; this combination may
be done equivalently by simultaneously fitting all three data sets
simultaneously, or by applying the joint posterior pdf from the X-ray
analysis as a prior for the SZ/lensing analysis, or vice versa. 
\begin{figure}
\centering\epsfig{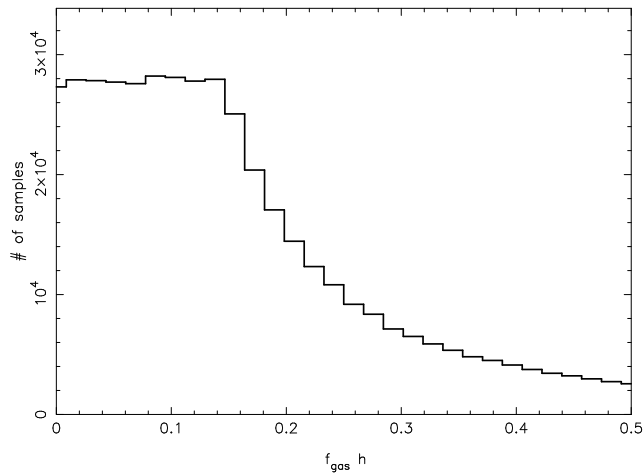}
\caption{ Implied prior probability distribution $\pr(f_{\rm gas}h)$,
as represented by an unsmoothed histogram of 50000 MCMC samples.}
\label{fig:fgasprior}
\end{figure}

In this work, we leave the $h$-dependence of the gas fraction as it is,
simply using the uniform priors of Table~\ref{tab:models} on 
$M_{\rm gas}$ and $M_{200}$. However, these apparently innocuous 
priors lead to an
informative prior on the combination~$f_{\rm gas} h$. This can be seen
by sampling the joint prior with no data using the MCMC 
algorithm~(Slosar \ea~\citeyear{COS/Slo++03}), and
computing $M_{\rm gas}/M_{200}$ for each sample in the same way as would
be done in the data analysis process. 
The resulting histogram is shown
in Figure~\ref{fig:fgasprior}. The effect of the 
upper limit on the gas mass can be seen as the turnover point at 
$M_{\rm gas, max}/M_{200, max} = 0.15$. 
Below this value,
the prior is uniform, whereas above it the larger values 
of~$f_{\rm gas} h$ are increasingly disfavoured, well reflecting our
prior knowledge of cluster masses.


\subsection{Current generation experiments}
\label{subsect:sim:current}

\begin{figure*}
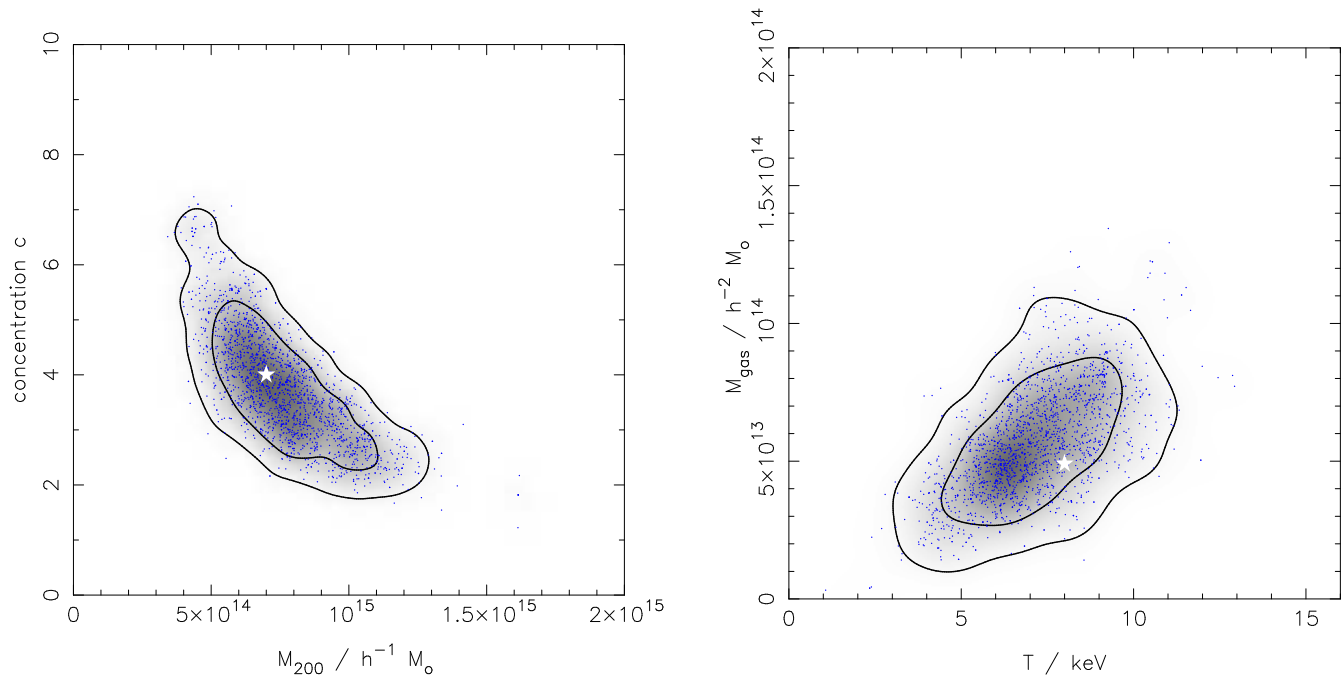

\begin{minipage}[b]{0.492\linewidth}
\centering\epsfig{file=current_iHSE-iHSE_001_hist2D_M200-c.ps,width=\linewidth,angle=0,clip=}
\end{minipage} \hfill
\begin{minipage}[b]{0.468\linewidth}
\centering\epsfig{file=current_iHSE-iHSE_001_hist2D_T-Mgas.ps,width=\linewidth,angle=0,clip=}
\end{minipage}
\caption{ Marginalised probability distributions of
cluster astrophysical parameters from mock current experiments.
Left: $\pr(M_{200},c|\Data,\imodel)$; 
Right: $\pr(T,M_{\rm gas}|\Data,\imodel)$. 
MCMC samples are shown as points, overlaid on their 
minimally-smoothed
histogram (marked with contours enclosing 68 and 90\% of the posterior
probability. The white star shows the true parameter values.}
\label{fig:VSA-pars}
\end{figure*}
\begin{figure*}
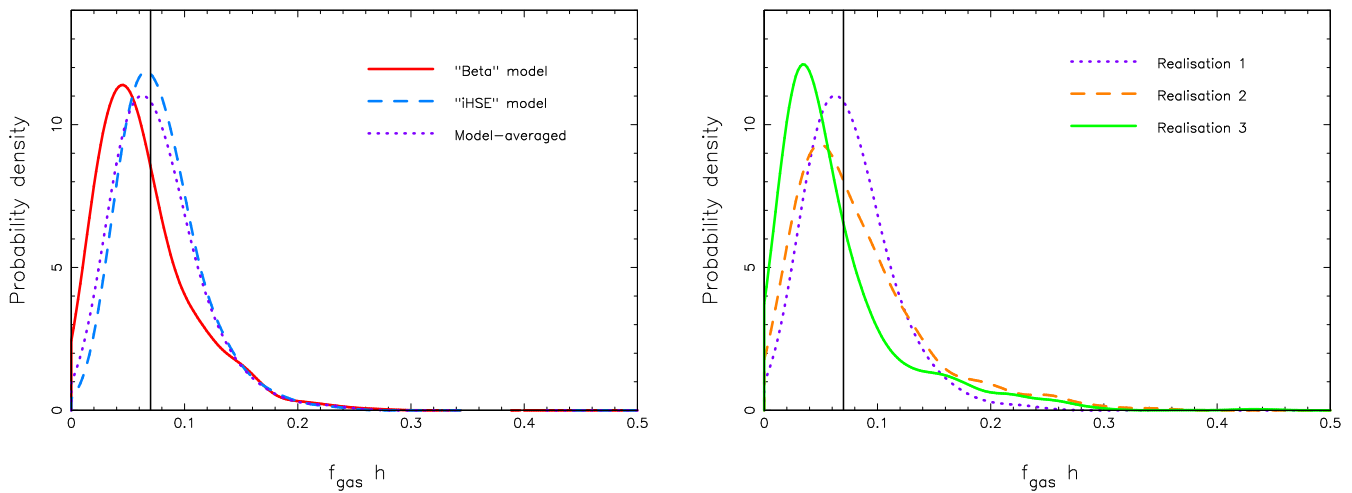

\begin{minipage}[b]{0.48\linewidth}
\centering\epsfig{file=current_iHSE_001_fgas.ps,width=\linewidth,angle=0,clip=}
\end{minipage} \hfill
\begin{minipage}[b]{0.48\linewidth}
\centering\epsfig{file=current_iHSE_all_fgas.ps,width=\linewidth,angle=0,clip=}
\end{minipage}
\caption{ Marginalised probability distributions of
cluster cosmological parameters from mock current experiments. 
Left: Marginalised probability distributions of
$f_{\rm gas}h$; the true cluster model is
iHSE.
$\pr(f_{\rm gas}h|\Data,\bmodel)$ and 
$\pr(f_{\rm gas}h|\Data,\imodel)$ are plotted with full and dashed lines
respectively;
the dotted curve shows the result of
model-averaging, $\pr(f_{\rm gas}h|\Data)$.
Right: the effect of different CMB realisations on the
model-averaged inferences. In both plots the true value of $f_{\rm
gas}h$ is shown by the dark solid vertical line. }
\label{fig:VSA-fgas}
\end{figure*}
Example model parameter inferences are shown in
Figure~\ref{fig:VSA-pars}, where in each panel
the posterior density has been
marginalised over all but two of the parameter dimensions. The effect of
the Gaussian prior on the gas temperature can be seen in the right-hand
panel. The estimated precision on each of the three parameters 
$M_{200}$, $c$, and~$M_{\rm gas}$ are 25, 25 and 50 percent 
respectively.  

The left-hand panel of Figure~\ref{fig:VSA-fgas} shows the cosmological
results from the joint
analysis of the current generation low redshift SZ and lensing data.
This plot shows the answer to question~(iii) as posed 
in Section~\ref{sect:infer}. The joint evidences calculated for each 
model,
for either given dataset, were equal within the numerical errors;
this indicates that the model averaging equation has two terms to be
considered, leading to the (quasi) model-independent statement about
$f_{\rm gas}$ given in the left panel of the figure. 

However, the right-hand panel shows the sensitivity of the VSA SZ data
to the contaminant 
primordial CMB fluctuations: the model-averaged $f_{\rm gas}$
probability distributions show significant variation with 
CMB realisation. 
Realisations 1, 2 and 3
correspond to the situations where the cluster lies approximately in 
front of a
primordial CMB saddle
point, a shallow trough and a peak respectively. 
In the latter case the gas mass
(and so gas fraction) is then underestimated. 

The evidence analysis of CMB realisations 2 and 3 show neither model
being preferred by more than a small factor in  probability ($<3$)
regardless of the true cluster model. In some cases the primordial CMB
is being fitted (by the more flexible Beta model) as well as the
cluster, and in others the noise is high enough for the Occam factor in
the evidence to dominate and the simpler iHSE model is preferred.
However, the extent to which the evidence favours either model is never
greater than the belief threshold suggested in
Section~\ref{sect:infer}.  This indicates that the presence of the
primordial CMB fluctuations has been dealt with correctly -- the
inconclusiveness of the evidence ratios ensure that  the conclusions
drawn about the structure of the cluster are not systematically
incorrect. 

However the implication of this result is
that in order to investigate the astrophysics of low redshift clusters 
via SZ and gravitational lensing (questions (i) and (ii)), more
information is
required. This could take the form of multi-frequency SZ observations,
to allow better separation of the cluster and primordial CMB
components~(Lancaster~\ea 2003, in preparation),
or stronger priors on the cluster parameters. Reducing the freedom of
the Beta model to fit the CMB fluctuations will indeed produce more
precise gas fraction estimates, but to be confident of the accuracy of
these numbers
the applied priors should be strongly physically motivated. 
A good first step in
this direction would be to derive joint priors on any model's parameters
from a large sample of hydrodynamically-simulated clusters.


\subsection{Next generation experiments}
\label{subsect:sim:future}

We now move on to consider the kind of observations we can expect from
upcoming SZ telescopes such as AMI~\citep{SZ/Kne++01}, the
SZA~\citep{SZ/Moh++02} and AMiBA~\citep{SZ/Lo++00}, in combination with
matched lensing observations from wide field optical cameras. Rather
than compare the different experiments, we note that they are
qualitatively similar instruments and proceed to use AMI, and for the
mock lensing observations the ESO Wide Field Imager, as specific
examples in this work. AMI will consist of 
(a) ten close-packed 3.7-m dishes operating at 15GHz
and (b) the eight 13-m dishes of the current Ryle Telescope 
(which are separated by longer baselines). The two parts of this
array will 
have different correlators and
therefore provide two independent measurements of the sky,
allowing the two datasets to be combined by a simple sum of the 
individual log-likelihoods.

An important question is that of the strength of the primordial CMB on
the angular scales to which AMI is sensitive. These correspond to a
maximum
$l$-range of 1000 to 6600 for AMI's small dishes, and 6000 to 36000 for
the large Ryle Telescope dishes; note that in practice the minimum
$\ell$~values used in observations will be significantly greater.
The primordial CMB power spectrum is
relatively poorly known in this region, with only the measurements by
ACBAR~\citep{COS/Run++03} extending to $\ell$~of 2500 and 
CBI~(Pearson \ea~\citeyear{COS/Pea++03}; Mason \ea~\citeyear{COS/Mas++03})  
extending to $\ell$~of 3500. The CBI group find an excess of power at 
the higher of these wavenumbers, which has been
interpreted as being due to the integrated SZ effect of
the large-scale structure along the line of sight~(Bond \ea~\citeyear{COS/Bon++03}).
For the purposes of this
work we make two assumptions: first, the primordial CMB can be
neglected for the Ryle observations at $\ell>6000$, and second that
extrapolating the power spectrum of Figure~\ref{fig:cmbspec}  to this
$\ell$ limit gives a reasonable estimate of the amplitude of the
primordial fluctuations on larger scales. For comparison with the latter
we also simulated short baseline data with no contribution to the noise
from the CMB. 
\begin{figure*}
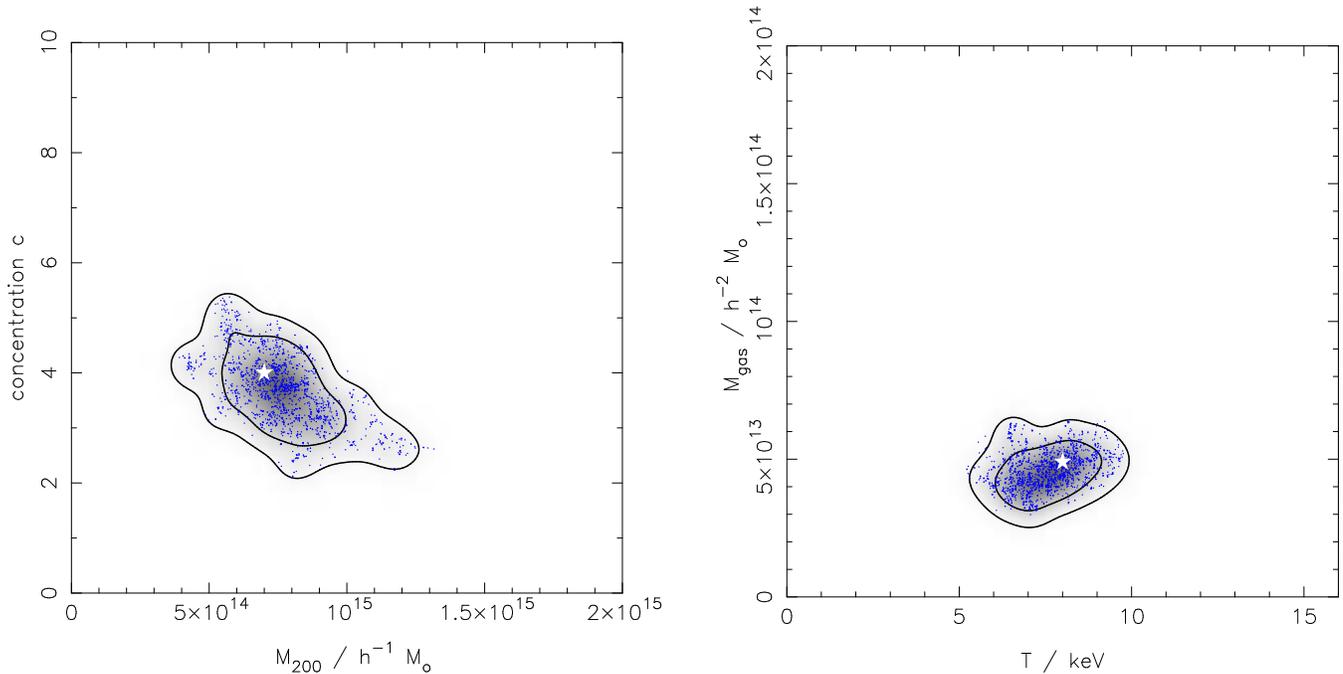

\begin{minipage}[b]{0.492\linewidth}
\centering\epsfig{file=nextgen_iHSE-iHSE_001_hist2D_M200-c.ps,width=\linewidth,angle=0,clip=}
\end{minipage} \hfill
\begin{minipage}[b]{0.468\linewidth}
\centering\epsfig{file=nextgen_iHSE-iHSE_001_hist2D_T-Mgas.ps,width=\linewidth,angle=0,clip=}
\end{minipage}
\caption{ Marginalised probability distributions of
cluster astrophysical parameters from mock next-generation experiments.
Left: $\pr(M_{200},c|\Data,\imodel)$; 
Right: $\pr(T,M_{\rm gas}|\Data,\imodel)$. 
MCMC samples are shown as points, overlaid on their 
minimally-smoothed
histogram (marked with contours enclosing 68 and 90\% of the posterior
probability. The white star shows the true parameter values.}
\label{fig:AMI-pars}
\end{figure*}
\begin{figure*}
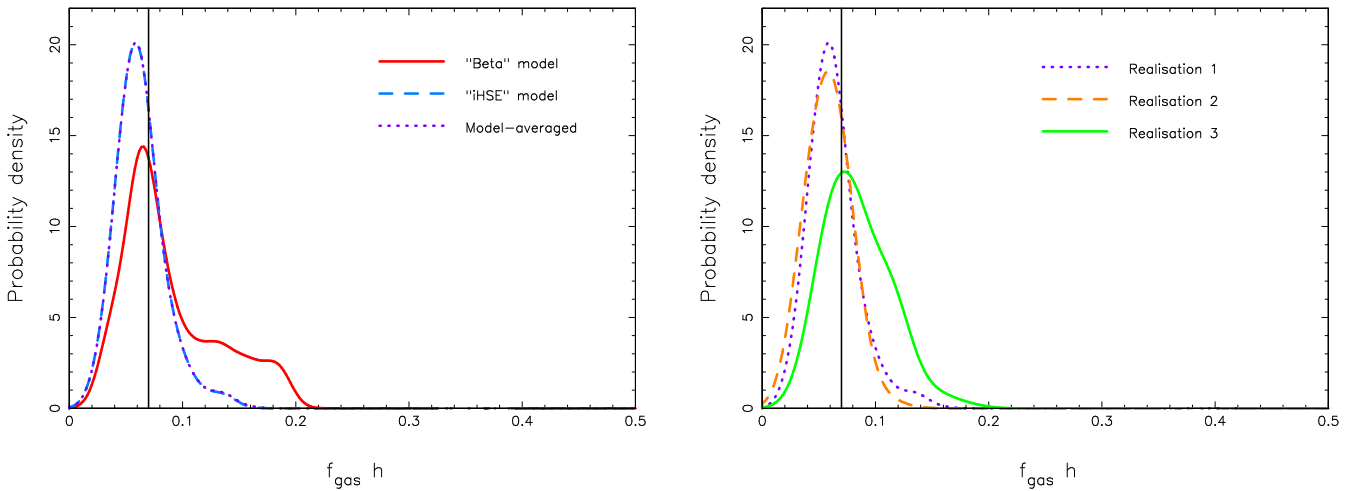

\begin{minipage}[b]{0.48\linewidth}
\centering\epsfig{file=nextgen_iHSE_001_fgas.ps,width=\linewidth,angle=0,clip=}
\end{minipage} \hfill
\begin{minipage}[b]{0.48\linewidth}
\centering\epsfig{file=nextgen_iHSE_all_fgas.ps,width=\linewidth,angle=0,clip=}
\end{minipage}
\caption{ Marginalised probability distributions of
cluster cosmological parameters from mock next-generation experiments. 
Left: Marginalised probability distributions of
$f_{\rm gas}h$; the true cluster model is
iHSE.
$\pr(f_{\rm gas}h|\Data,\bmodel)$ and 
$\pr(f_{\rm gas}h|\Data,\imodel)$ are plotted with solid and dashed
lines respectively; the dotted curve shows the result of
model-averaging, $\pr(f_{\rm gas}h|\Data)$, and is indistinguishable
from the (correctly) selected model curve. Right: the effect of 
different CMB realisations on the model-averaged inferences. 
In both plots the true value of $f_{\rm gas}h$ is shown by the 
dark solid vertical line. }
\label{fig:AMI-fgas}
\end{figure*}

Example model parameter inferences from this quality data are shown in
Figure~\ref{fig:AMI-pars}, where in each panel 
the posterior density has been
marginalised over all but two of the parameter dimensions. The effect of
the Gaussian prior on the gas temperature can be seen in the right-hand
panel. The estimated model-dependent precisions on each of the three 
parameters $M_{200},
c$ and~$M_{\rm gas}$ are 20, 20, and 12 percent respectively. The
improvement in the results from the lensing data is slight, the
increased lensing strength being balanced by the decreased number of
observed background galaxies due to the smaller field of view.
The improvement in gas mass estimation is
more marked, due to a combination of reduced primordial CMB at the
higher $l$-values, and the more comprehensive $uv$-plane coverage
afforded by AMI. 

\begin{table}
\caption{Evidence ratios from the joint analysis of the 
``next-generation'' mock observations; this quantity is given by
$\pr(\Data|\textrm{correct model})/\pr(\Data|\textrm{incorrect model})$ 
and so should be greater than unity if accurate astrophysical
conclusions are to be drawn.} 
\label{tab:AMIevid}
\begin{center}
\begin{tabular}{ccc}
CMB included?	& Model 	& Evidence ratio (CMB real$^{\rm n.}$ 1, 2, 3)  \\ \hline\hline
Yes			& Beta	& 1.5, 81, 2.5			\\
			& iHSE	& 110, 110, 37			\\ \hline
No			& Beta	& 20			\\
			& iHSE	& 600			\\ \hline
\end{tabular}
\end{center}
\end{table}

Table~\ref{tab:AMIevid} gives the evidence ratios for the experiments
outline above. In the case where no contaminant primordial CMB is
present the most probable model matches that used in simulating the
data. The factor by which the true model is more likely is larger when
the iHSE model is used: this model is both simpler, and it provides a 
better fit. The same is
true when the primordial fluctuations are present, with suitably (but
not greatly) reduced
evidence ratios, with the variation in the evidence ratios being due to
differing noise
realisations. 
We again see the sensitivity to the noise realisation, with confident
model
selection possible in only 4 out of the 6 analyses. 
For the Beta model cluster, the same balance
between model complexity and goodness of fit is seen as in 
Section~\ref{subsect:sim:current}, suggesting the need for informative 
priors even for this higher quality data.

Figure~\ref{fig:AMI-fgas} shows the cosmological inference drawn from
the analysis of the mock iHSE cluster. The model-averaged 
$f_{\rm gas} h$
probability distributions are dominated by the iHSE model contribution
-- this astrophysical model has been successfully selected by the 
evidence.  
Taking the median sample as an estimator for the gas fraction we find 
$f_{\rm gas}h=(0.061^{+ 0.019}_{- 0.015})$. With no CMB contamination
this estimate changes to $f_{\rm gas}h=(0.054^{+ 0.015}_{- 0.013})$,
an increase
in precision of just 5\% (from 28 to 23\%). This is an indication of the
small contribution to the error budget that the primordial CMB has at
these angular scales.
Indeed, as previously 
mentioned the shortest AMI baseline will be longer than that used in
this work, such that 
the effect of the primordial CMB will be reduced; we might therefore
expect results
lying inbetween the two situations simulated here.


\section{Discussion}
\label{sect:disc}

The analysis of the simulated data presented in the previous section was
designed to be a simple demonstration of a general methodology. In the
current section we discuss the advantages and disadvantages of our
approach, and its ability to be extended, beginning with some comparison
with other methods currently in use. 

The Bayesian method described here can be seen as a generalisation of
the model fitting procedures employed by other workers. For example,
\citet{GL/KCS02} fitted lens models with 1, 2 and 3 parameters to weak
shear data for Abell~1689 by the maximum likelihood method; they 
compute likelihood contours for the parameter uncertainties and compare
the models' goodness of fit with the likelihood ratio test. As
explained in Section~\ref{sect:infer}, such a grid-based computation is
not practical with the 6--8 parameters used here. Indeed, numerical
maximisation of a function of 6--8 variables is already a demanding
problem, especially when there are multiple maxima to be investigated. 
Comparing models by their maximum likelihoods is also rather sensitive 
to noise features in the data, rather than assessing the relative 
appropriateness of the models to the task of explaining the data. The
maximum likelihood ratio is formally  equivalent to the evidence ratio
when the prior pdf is a delta-function at the best-fit point: this is
clearly not an accurate representation of our prior knowledge. The
$\chi^2$ fitting of SZ visibility data by \citet{SZ/Gre++01}, and the 
joint maximum likelihood analysis of SZ and X-ray data of
\citet{J/Ree++02} are two other examples of a  small number of
parameters being fitted within the context of a single cluster model;
one of the aims of this work was to provide a complete framework which
combined and extended analyses such as these.

Other methods proposed for use in the joint analysis of cluster data
have, to date, been focused on ``parameter--free'' reconstruction.
\citet{J/Reb00} suggests an iterative procedure for refining the
cluster potential using X-ray, SZ and weak lensing images, whilst
\citet{J/Zar++01} provide a direct inversion method to take these
images and produce a three-dimensional cluster model, making use of
some attractive features of working in the Fourier domain.
\citet{J/Dor++01} suggest using SZ and weak lensing data to constrain
successive orders of perturbation from spherical symmetry, again
working from ready-made maps. All these methods have been shown to work
well with noiseless input. However, those working with the observations
have opted to increase the complexity of their cluster models in a more
gradual way, for example fitting an X-ray map with a simple model and
using this model to predict the observed SZ
effect~\citep[\eg][]{J/Jon++01}. The method described here is this
common sense reduced to calculation -- the use of simple functions
for the
potential, gas density and temperature allows the model complexity to
be tuned to the data quality via the evidence. The ``parameter-free''
methods referred to above actually have many parameters, usually the
values of pixels in a grid: direct methods will produce one set of
parameters, but these may not be the most probable given the data, or
even the most appropriate given the parameter degeneracies. When trying
to measure quantities such as the gas fraction, all the cluster
configurations allowed by the data should be accounted for, in order to
calculate an accurate confidence interval; only by fully 
exploring a model's parameter space this can be achieved. Indeed,
consideration of a range of models is then desirable to gain the next
level of accuracy, one that is model-independent in the sense of the
discussion in Section~\ref{sect:infer}.
Sampling from a model's parameter space produces the set of 
cluster configurations permitted by the data, 
which is  arguably more useful than the unique
solutions generated by direct methods.

One might wonder how the MCMC technique endorsed in this work would
perform if used in the many-parameter modelling of the type mentioned
above. Indeed, in this context the number of parameters included here 
is rather small.  The evidence itself is a guide in the development of
these methods, providing a handle on the information content of the
data: if the evidence does not favour a triaxial ellipsoid over a
spherical model, then it might reasonably be assumed not to
 favour a ``parameter-free''
representation. The ability of a sampler to cope with increasing
numbers of parameters is somewhat sensitive to the shape of the
posterior distribution under investigation; we find that including
extra nuisance parameters (the parameters of  point sources
contaminating the SZ data for example) does not affect the accuracy of
the posterior  exploration~(Lancaster~\ea 2003 in preparation), and
neither
does increasing the number of sub-clumps when modelling gravitational
lenses with multiple mass concentrations~\citep{GL/Kne++03}. Moving to
three-dimensional cluster modelling introduces a number of strong
degeneracies in the parameter space~\citep{J/F+P02} which may well
require a tailor-made MCMC sampler instead of the general purpose
engine used here.  Such a sampler might be expected to reduce the
run-time of the method, which (compared to a direct inversion or a
downhill simplex $\chi^2$  minimisation) is its major disadvantage: to
produce each posterior distribution shown in the figures of this paper
a computation time of several hours with a 1~GHz processor was required.
The evidence calculation takes rather longer, since the numerical
precision comes from repeated posterior explorations. The computation
time scales approximately as $N_{\rm data} \times
N_{\rm parameters}$~\citep{MCMC}, prompting careful design of the sampler
and likelihood calculation. However, the alternatives for coping with
noisy data can be just as time consuming; for instance resampling of
the data to generate confidence limits~\citep[\eg][]{COS/All++03} 
effectively performs the same calculations as the MCMC process, but
with more limited output. Nevertheless, it is the computational cost of
the method that is perhaps the most urgent aspect to be addressed in
further work.


\section{Conclusions}
\label{sect:concl}

We have developed an algorithm based on the Markov-Chain Monte-Carlo
technique for investigating simple but many-parameter models of
clusters. The method allows straightforward inclusion of many
datasets, correctly weighting each datapoint according to its assumed
likelihood; by exploring the posterior probability distribution rather
than the likelihood we can incorporate information on the cluster from
other sources via the parameter prior densities. Calculation of the
Bayesian evidence by thermodynamic integration during the burn-in period
can be done to sufficient accuracy to allow different astrophysical
models to be compared; this statistic automatically includes the
common-sense of Occam's razor, allowing movement away from the simplest
assumed models only when the data require it.

Applying the method to
simulated weak gravitational lensing and interferometric 
Sunyaev-Zel'dovich effect data we draw the following conclusions:
\begin{itemize}
\item Gravitational weak lensing data allow cluster mass distribution 
model parameters to be estimated with a precision of around 20 percent
over the small range of low redshifts discussed here.
\item 
Primordial CMB anisotropies contaminate SZ
observations on angular scales less than $l\approx1000$. 
However, the nature of the primordial fluctuations allows them to be
properly accounted for in the likelihood function, preventing 
inaccurate conclusions about either the cluster model or the gas
fraction.
The available 
precision on the cluster gas mass
increases by a factor of approximately two (to around 10
percent) when moving from
instruments such as the VSA to those more like AMI. 
\item The behaviour of the Bayesian evidence as a model selection tool
can be understood in terms of both goodness of fit and model complexity;
in the case of the primordial CMB contaminated data, the more flexible
fitting formulae are sometimes favoured by the evidence as they provide
a better
fit to all the data, prompting the need for improved prior constraints
on the model parameters. One recommended source of these priors is a
sample of numerically simulated clusters.
\item Where the SZ data are less contaminated by the primordial CMB the
evidence does indeed allow successful astrophysical model selection,
leading to accurate conclusions about the dynamical state of the
cluster under observation.
\item Where the evidences for a range of models are of comparable size,
a correctly weighted average may be taken, resulting in appropriately
precise inferred uncertainties on the parameter in question -- for the
gas fraction $f_{\rm gas}h$ we may expect a model-independent
uncertainty on this parameter of around 50 percent for a low redshift
cluster, and under 30 percent with observations of a cluster with AMI.
\end{itemize}

This last point is one worth returning to; under the assumption of a
Universal cluster gas fraction the model-independent inference for each
independent member of a sample of  $N$~clusters  can be combined by
straightforward multiplication of their posterior probability
distributions, reducing the uncertainty on this parameter
by a factor of approximately~$\sqrt N$. 

The work described in this paper is straightforwardly extendable to
incorporate X-ray, and indeed any other, cluster data. 
Similarly, investigating more
complex cluster models by relaxing the assumptions of isothermality,
spherical symmetry and a single cluster potential is easy to
do, with the evidence providing the self-consistent and logical way
through the astrophysical model analysis. The obvious immediate next 
step to take is the inclusion of the X-ray data, and the opening up of
another dimension of cosmological parameter space, that of the Hubble
parameter: this analysis applied to clusters observed with the VSA will
be the subject of future publications.


\subsection*{Acknowledgments}

We thank Sarah Bridle, Keith Grainge, Steve Gull, Katy Lancaster,
Charles McLachlan and Richard Saunders for helpful discussions, and are
very grateful to John Skilling of MaxEnt Data Consultants for making
his MCMC software \texttt{BayeSys3} available to us.  
We thank the anonymous referee for suggestions that improved the
readability of the article.
PJM acknowledges
the PPARC, and AS acknowledges St.\ John's College,  Cambridge, for
financial support in the form of research studentships.

\bibliographystyle{mn2e}
\bibliography{MD680rv}


\bsp 
\label{lastpage}
\end{document}